\documentstyle[times,pramana,epsf,floats]{ias}
\newcommand{\cp}{{\cal P}}
\newcommand{\be}{\begin{eqnarray}}
\newcommand{\ee}{\end{eqnarray}}

\begin{document}
\mark{{What's new at small x}{Raju Venugopalan}} \title{What's 
new at small x\footnote{Invited talk at the 6th Workshop on
    High Energy Particle Physics, Jan. 3-15 2000, Chennai, India}}

\author{Raju Venugopalan} 
\address{Physics Dept.,Brookhaven National Laboratory,
  Bldg 510A, Upton, NY-11973, USA} 

\abstract{We discuss some recent developments in small x physics.}

\maketitle
\section{Introduction}
\vskip 0.05in

In the last couple of decades, following the discovery of asymptotic
freedom in the early 70's, perturbative QCD has been enormously
successful in describing the physics of very high $Q^2\gg
\Lambda_{QCD}^2$.  However, it is sobering to think that these
very high $Q^2$ processes comprise the tails of distributions--their
contribution is a small fraction of the total cross--section at high
energies.  The vast bulk of the cross--section, corresponding to soft
and semi--hard processes, is still ill understood in QCD. Note that
for momentum transfer square $Q^2$, the high energy
limit $s\longrightarrow \infty$ is also the limit $x\longrightarrow 0$
since $x\propto Q^2/s$. The physics of high energy soft and semi--hard
processes in QCD is therefore also the physics of small
$x$. Understanding the origin of these small $x$ processes within the
framework of QCD is an outstanding challenge to both theory and
experiment. In this talk, I will attempt to summarize some recent
theoretical studies on the physics of high energy (or small
$x$) processes in QCD. Note: Due to space--time limitations, several important 
topics will not be treated--vector meson production, nuclear shadowing, etc. 
Also, given the large amount of activity in the 
field, it is unavoidable that my survey of the literature will not be 
as complete as it should be.

Phenomenological ideas motivated by Regge theory have 
had some success in describing some of the data. For example, the 
t--channel exchange of an object with vacuum quantum numbers, the 
notorious Pomeron, provides a reasonable description of total cross--sections 
at very high energies~\cite{DonnLndshff}. It corresponds to a 
simple pole, with unit intercept, for the amplitude 
in the $(t,j)$ plane, which through a Sommerfeld--Watson transformation to 
the $(s,t)$ plane gives cross--sections rising with the energy:
\be
A(t,j)\sim {1\over {j-\alpha(t)}}\,\,\,\stackrel{\rm SW}{\longrightarrow} 
\,\,\,A(s,t)\sim s^{\alpha(t)} \, ,
\label{pomamp}
\ee
where $\alpha(t) = \alpha_0 + \alpha^\prime t$. For the Pomeron, 
phenomenologically, $\alpha_0\approx 1.08$ and the string tension 
$\alpha^\prime = 0.25$ GeV$^{-2}$.
Invoking the optical theorem, one can easily show that 
\be
\sigma_{{\rm tot}} \sim s^{\alpha(0)-1} \, .
\ee
Donnachie and Landshoff~\cite{DonnLndshff} have shown that 
$\sigma(s)\sim s^{0.08}$ provides a good fit to the available 
data on $\bar{p}p$, $pp$, $\pi p$, $Kp$ and $\gamma p$ collisions at high 
energies. 
Strictly speaking, they find that 
\be
\sigma(s) = A s^{0.08} + B s^{-0.45} \, ,
\ee 
gives a good fit to the above mentioned data. The term with the decreasing 
contribution as a function of the energy corresponds to the ``Reggeon'' 
exchange of $\rho$,$\omega$,$f_2$, and $a_2$ mesons. The Regge form of the 
amplitude also explains high energy, small $|t|$, differential cross--sections 
and the shrinkage of the diffractive peak at high energies.

Despite the apparent phenomenological successes of the Pomeron
concept, we still don't have a very good idea of what it is or why it 
works. A popular
conjecture is the one first postulated by Low and by 
Nussinov~\cite{LowNussinov}, where Pomeron exchange is taken to be the
color singlet component of two gluon exchange in the t--channel. In
weak coupling, Lipatov and collaborators have shown that the leading
logarithmic $\alpha_S\ln(1/x)$ result in perturbative QCD corresponds to the
$t$--channel exchange of two ``reggeized'' gluons. In the color
singlet channel, they constitute the well known BFKL hard
pomeron~\cite{BFKL}. Recent developments suggest however that this
Pomeron may not be entirely robust at next to leading order, thereby 
confounding Pomeron enthusiasts. A possible resolution within
the Pomeron framework is that multi--Pomeron exchanges become
important sooner than one expects them to.   
I say sooner because one
expects these exchanges to become important eventually anyway since
they help ensure that unitarity is satisfied at asymptotically high
energies. The Onium--model of Mueller~\cite{Mueller94}, 
where the mass of the Onium 
pair provides a perturbative scale, provides the framework of 
several recent multi--Pomeron studies. See for instance Ref.~\cite{Kovchegov}.

Alternatively, one may eschew the Pomeron language altogether. 
One such approach describes the physics of high density QCD within a 
Wilsonian renormalization group improved~\cite{JKMW,JKW} 
classical effective field theory (EFT) of small $x$ QCD~\cite{MV}. 
One may expect that the two approaches are related. Recently, it has  
been argued that the formalism of multi--Pomeron exchanges 
can be recovered as a limit of the Wilsonian renormalization group 
formalism~\cite{KMW}.

With the advent of HERA in the early 90's, one was able to explore 
the regime in QCD where $x=Q^2/s\ll 1$ but $Q^2\gg \Lambda_{QCD}^2$. The 
advantage of this regime is that while the coupling $\alpha_S(Q^2)\ll 1$, 
large logarithms $\alpha_S\ln(1/x)\sim 1$ make the physics non--perturbative. 
From a theoretical point of view, this regime of high parton densities is 
interesting since it affords one the opportunity to study the interplay 
between perturbative and non--perturbative physics. In particular, one 
expects to see the effects of the high parton densities we mentioned 
above. An especially useful 
probe of this interplay is hard diffraction, which constitutes a significant 
part of the cross--section in deeply inelastic scattering (DIS). 
In this case, there is a 
color singlet exchange between the hadron and the projectile, the latter 
fragmenting into a hard final state. Interestingly, one can show that 
the usual factorization theorems apply to diffractive DIS~\cite{Collins,CFS}. 
We will discuss diffraction further later on in this talk.

This talk is organized as follows. In the next section, we briefly 
discuss the DGLAP double log limit at small $x$. We will comment on how 
well it does in describing the HERA data. In section 3, we 
will discuss the BFKL equation, and the next to leading order BFKL equation. 
Next, 
we will discuss different approaches to the physics of high parton 
densities, which study the regime where the linear evolution equations in 
$x$ and $Q^2$ break down. In section 5, we discuss recent results on hard 
diffraction. In Section 6, we make the connection between small $x$ physics 
and heavy ion collisions. We stress the importance of understanding the 
small $x$ component of the nuclear wavefunction to better understanding 
the initial conditions and possible thermalization in heavy ion collisions 
at RHIC and LHC. We end with a brief outlook on future directions in 
theoretical and experimental studies of high energy QCD.

\section{DGLAP evolution in QCD}
\vskip 0.05in

This topic is of course highly developed and discussed at length 
in several textbooks. We will discuss particular aspects of it in order 
to motivate the discussion in following sections. This section and 
the next have been influenced in part by the nice lectures of Salam--for more 
details, we refer the reader to them~\cite{Gavin}.
Let us begin with the canonical process, DIS of 
electrons off hadrons or nuclei. The kinematic invariants here are 
\be
x_{Bj} = {-q^2\over 2P\cdot q}\,\,;\,\, Q^2=-q^2> 0\,\,;\,\, y ={P\cdot q
\over P\cdot k}\,\,;\,\, s = 2P\cdot k \, ,
\ee
and these satisfy the relation $xy=Q^2/s$. 
In the rest frame of the target, the 
virtual photon fluctuates into a quark anti--quark pair, which subsequently 
interacts with the target. The $q\bar{q}$ pair undergoes Bremsstrahlung, 
emits a gluon, which subsequently splits into another (or with differing 
probability a $q\bar{q}$ pair), and so on until the parton on the lowest 
rung of the ladder interacts with the target. Each rung of the ladder 
therefore contributes a Bremsstrahlung phase space integral 
\be
\alpha_S\int\,{d^2 k_t\over k_t^2}\int\,{dx\over x}\,\,\,\longrightarrow
\,\,\,\alpha_S^p\, \ln^q \left({x_0\over x}\right)\, 
\ln^r \left({Q^2\over Q_0^2}\right) \, .
\ee
If we  are interested in the kinematic region $x_0\sim x_{Bj}$, and 
$Q_0^2\ll Q^2$, clearly the dominant contribution to the scattering will 
be  $(\alpha_S\ln({Q^2\over Q_0^2}))^n$ logs which are summed over all $n$. 
The DGLAP equations are renormalization group (RG) equations that sum up these
large logarithms~\cite{DGLAP}.

In at least a good chunk of the region probed by HERA, the kinematics
are such that it is likely that the leading contribution is from logs
in both $x$ and $Q^2$--one therefore sums large logs in both $x$ and
$Q^2$: $(\alpha_S\ln(x_0/x)\ln(Q^2/Q_0^2))^n$. The unintegrated gluon
distribution $G(x,Q^2)=d(xg(x,Q^2)/dQ^2$ satisfies the RG--equation
\be
G(x,Q^2)=G^{(0)}+\int_x^1 {dz\over z}\,\int\, dk_t^2
\, K_{DGLAP}(Q^2,k_t^2)\, G\left({x\over z},k_t^2\right)\, ,
\label{inteqn}
\ee
where $K_{DGLAP}(Q^2,k_t^2)= {\bar\alpha}_S\theta(Q^2-k_t^2)/Q^2$, and 
${\bar\alpha}_S=\alpha_S N_c/\pi$.
This integral equation is diagonalized by the simultaneous Mellin transform 
with respect to $x$ and $Q^2$. For a particular initial condition, 
\be
{\tilde G}_{\gamma,\omega} = {1\over \gamma}\,{1\over (\omega-{\alpha_S\over\gamma})}
\, ,
\ee
where $\gamma$ is the leading order gluon anomalous dimension. 

First performing the inverse Mellin 
transform with respect to $\omega$, and then 
performing a saddle point integration 
over $\gamma$, one obtains the well known result~\cite{GrossRujula}
\be
Q^2\,G(x,Q^2)\approx {1\over 2}
\left({1\over \pi^2 \alpha_S\ln(1/x)\ln(Q^2/Q_0^2)}\right)^{1\over 4}
\,
\exp\left(2\sqrt{\alpha_S\,\ln({1/x})\,
\ln({Q^2/Q_0^2})}\,\right) \, .
\ee
The strong rise in the gluon distribution is mirrored by the structure 
function,
\be
R_F\,F_2(x,Q^2) = \exp\left(\kappa\right) \, ,
\ee
where $R_F$ is a coefficient which also depends on $x$ and $Q^2$, 
and $\kappa=\sqrt{\log(1/x)\log(Q^2/Q_0^2)}$. 

This is the so--called double asymptotic scaling, the slope of the 
structure function is a universal quantity. In this
region, the scaling violations are relatively independent of the
particular form of input parton distributions. At HERA, the running
coupling has been extracted in the double log region: a NLO fit gives
$\alpha_S(M_Z)$ = 0.120$\pm$ 0.05(exp.)$\pm$ 0.09
(theory)~\cite{BallForte}.  The theoretical uncertainities include
estimates of small $x$ corrections to the NLO calculation, and
factorization and scale uncertainities~\cite{Forte}.  There are also
some unresolved issues regarding the normalization $R_F$ of double
asymptotic scaling in the HERA kinematic region~\cite{ManSaaWei}.
A potential problem with this nice picture is the possibility that the
contribution of higher order corrections to the DGLAP gluon anomalous
dimensions (in the same kinematic region) induce an even faster
rise--opening a Pandora's box of problems\cite{Forte}.  These will be
directly related to the issues discussed in the following sections.

\section{The BFKL and NLO BFKL summations}
\vskip 0.05in

The BFKL summation corresponding to summing only leading logarithms in 
$x$--$(\alpha_S\ln(1/x))^n$, is applicable when there are two large scales 
in the problem $Q^2\sim Q_0^2\gg \Lambda_{QCD}^2$. Consider again the 
integral equation discussed in Eq.~\ref{inteqn}. The BFKL equation is 
obtained by replacing $K_{DGLAP}\longrightarrow K_{BFKL}$, where the 
BFKL kernel is~\cite{BFKL}
\be
K_{BFKL} = {\bar\alpha}_S \,\left({1\over |\vec{Q}-
\vec{k_t}|^2}-\delta(Q^2-k_t^2)\,\int^{k_t} {d^2 p_t\over {\pi p_t^2}}
\right) \, .
\ee
Again, as discussed in the previous section, the integral equation can 
be solved by performing the Mellin transform, which reads
\be
{\tilde G}_{\gamma,\omega} = {\omega {\tilde G}_{\gamma,\omega}^{(0)}
\over (\omega-{\bar\alpha}_S\chi(\gamma))} \, .
\ee
The function $\chi(\gamma)$, defined to be $\chi(\gamma)=
\psi(1)-0.5\cdot(\psi(\gamma)+\psi(1-\gamma))$, 
(where $\psi$ is the logarithmic derivative of the gamma function) is known 
as the characteristic function.

Taking the inverse Mellin transform with 
respect to $\omega$, and performing a saddle point expansion of the 
$\gamma$ integral around $\gamma=1/2$, one obtains the result
\be
G(x,Q^2)\approx {x^{-{\bar\alpha}_S\chi(1/2)}\over \sqrt{2\pi {\bar\alpha}_S
 \chi^{\prime\prime}(1/2)\log(1/x)}}\, {1\over Q Q_0} \, ,
\ee
where $\chi^{\prime\prime}$ is the second derivative of $\chi$ with respect 
to $\gamma$. For $\alpha_S=0.2$, the power of $x$ is $-{\bar\alpha}_S\chi(1/2) 
=-0.5$, a rise that's too rapid to be compatible with the HERA data. 
Furthermore, the scaling violations are also incompatible with the HERA 
data. Argueably, since the BFKL equation is derived under the assumption 
that $Q^2\sim Q_0^2\gg \Lambda_{QCD}^2$, one shouldn't expect it to 
explain the HERA DIS data. BFKL--like effects have been studied, for instance, 
in $\gamma^{*}-\gamma^{*}$ scattering at LEP, and in jet production at 
HERA and Fermilab.

Of greater concern, conceptually, is the fact 
that the solution to the BFKL equation exhibits $k_t$ 
diffusion~\cite{McDermott}. Although the typical momenta in the BFKL ladder 
are hard, the solution ``diffuses'' to the infrared at small $x$--the solution 
is therefore sensitive to momenta in the non--perturbative region.

Until recently, it was believed that next-to-leading-log (NLL) resummation of 
the form $\alpha_S (\alpha_S\ln(1/x))^n$ might clarify the theoretical 
picture.  In the Mellin transform language, the characteristic function 
$\chi$ can be expanded as 
\be
{\bar \alpha}_S\chi(\gamma) = {\bar\alpha}_S\chi_0(\gamma) + 
{\bar \alpha}_S^2 \chi_1(\gamma) + O({\bar\alpha}_S^3) \, ,
\ee 
where $\chi_0$ is the usual BFKL characteristic function we discussed above, 
and $\chi_1$ is the NLL-term. The computation of this term took about 
10 years (!) and was done independently by two groups~\cite{FLCC}.
The result is 
\be
\chi\left({1\over 2}\right) = \chi_0\left({1\over 2}\right)\, 
\left[1-6.47{\bar \alpha}_S\right]\, .
\ee
The power of the gluon distribution then for ${\bar \alpha}_S=0.2$ is 
$-0.15$. The correction is thus not only large, it also switches sign! 
Also, the structure of $\chi(\gamma)$ is now very different. One now has 
complex saddle points which give rise to cross--sections which, albeit 
real, oscillate with $\ln(Q^2/Q_0^2)$~\cite{Ross}.

Clearly, the resummation procedure, as developed thus far, is flawed. There 
have been several suggestions recently on how one may ``cure'' this 
result. One detailed proposal~\cite{CCS} suggests that even though 
the full NLL characteristic function $\chi_1$ has many contributions, 
a few collinear contributions give the bulk of the contribution. There 
are collinear corrections arising from a) running coupling effects, 
b) the non singular part of the splitting functions, and c) the choice 
of energy scale. Collecting these, one obtains the relatively 
simple collinear contribution to the NLL characteristic function
\be
\chi_1^{coll}(\gamma)= {A_1\over \gamma^2}+ {(A_1-b)\over (1-\gamma)^2}
-{1\over 2\gamma^3} -{1\over 2(1-\gamma)^3} \, ,
\ee
where $A_1=-11/12$, and $b=11/12-n_f/6$ . The authors of Ref.~\cite{CCS} 
have shown that $\chi_1^{coll}$ is in very good numerical agreement with 
the full NLL result. These collinear contributions can now be summed to 
all orders, and give rise to stable results (as a function of $\alpha_S$)
for the gluon anomalous dimensions, and for the exponent of the gluon 
Green's function at high energies.

The specific proposal we discussed briefly is very elegant and clever. It 
relies heavily though on the idea of ``collinear dominance'' to all orders. 
Whether this is indeed the case is not entirely clear at present. For 
instance, multi--pomeron (or high parton density) effects become 
important~\cite{Mueller97} 
at rapidities $y_{mult} \sim {1\over (\alpha_P -1)}\ln(1/\alpha_S^2)$, 
where $\alpha_P-1=4{\bar \alpha}_S\ln(2)$. However, 
certain running coupling effects become important~\cite{KovMuell} at 
rapidities $y_{NLO}\sim 1/\alpha_S^{5/3}$. Thus, parametrically, 
multi--Pomeron  effects appear sooner than some 
running coupling effects. This fact 
is not taken into account in BFKL--related proposals.

\section{Classical EFT and ``Onium'' approaches to high parton densities}
\vskip 0.05in

As one goes to higher energies, smaller $x$'s, 
one might ask whether there is a simpler organizing principle than 
computing an endless number of diagrams. For example, the properties of 
condensed matter systems in the vicinity of a critical point can be 
formulated in terms of effective theories which capture much of the 
physics. Attempting to compute critical behavior in the full theory would 
be an impossibly difficult task.

In small $x$ physics, our quest for the right effective theory is helped 
by the following. Firstly, since the density of partons is growing with 
energy, occupation numbers become large. This makes it likely that 
classical methods are applicable. Secondly, at high 
energies, in the infinite momentum frame,  partons have large field 
strengths on the transverse sheet-corresponding to a large parton 
density per unit area. This provides a scale, which at sufficiently 
small $x$, is large enough to make weak coupling methods feasible.
Finally, since small $x$ partons are short-lived relative to partons at 
large $x$, the latter act as static sources whose dynamics can be ignored, 
a la Born--Oppenheimer, on the time scales of interest.

Rather ironically, the problem is simpler to formulate for a large
nucleus~\cite{MV}, where there are $A^{1/3}$ more partons per unit
area on the transverse sheet than in a hadron. Since the sources are
confined in different nucleons, they are uncorrelated. Classical
parton distributions can then be computed as correlation functions of
a 2--dimensional Euclidean field theory with random, Gaussian
sources. The problem is then formally just like that of computing the
infra-red properties of a spin glass~\cite{RajGavai}. In practice,
path ordering of the space--time rapidity is necessary, and leads to
an analytical solution for classical parton distributions in the
``Colored Glass Condensate''~\cite{JKMW,Kov}. At transverse momenta
$k_t\gg Q_s$, where $Q_s\gg \Lambda_{QCD}$ is a saturation scale,
parton distributions have the usual Weizs\"acker--Williams $1/k_t^2$
behavior. However, for $k_t\leq Q_s$, their behavior saturates,
growing only logarithmically at small $k_t$.

Quantum corrections to the classical EFT give large 
logs in $\alpha_S\ln(1/x)$~\cite{AJMV}. A Wilson renormalization group 
procedure was devised which sums up these large logs~\cite{JKMW}. The 
form of the effective action remains the same as one goes to small $x$--
the only thing that changes is the weight function for the sources. 
For Gaussian sources, this gives $Q_s\longrightarrow Q_s(x,Q^2)$. In general, 
the weight function obeys a non--linear Wilson renormalization group 
equation~\cite{JKW}. In the limit of low parton densities, it is just 
the BFKL equation we discussed previously~\cite{JKLW1}. At large $Q^2$, an 
all twist result is obtained~\cite{JKLW2}, whose leading term is the 
small $x$ DGLAP equation, and the next-to-leading term is the higher 
twist correction previously computed by Gribov, Levin, and Ryskin~\cite{GLR}, 
and by Mueller and Qiu~\cite{MuellQiu}. The outstanding question is whether 
this approach, by incorporating high parton density (multi--ladder) 
effects already at leading order through the non--linearities of the 
classical field, provides a more stable expansion 
than the BFKL--motivated approach.

One can also compute the structure function $F_2$ to all orders in the 
classical background field~\cite{MV99}. For Gaussian sources, one recovers 
the Glauber formula~\cite{M90NZ} originally 
derived in the nuclear rest frame
\be
F_2 &=& {Q^2 N_c\over (2\pi)^3}\int d^2 b \int_0^1 dz\,
\int_0^{1\over \Lambda_{QCD}} dx_t x_t \,\left(1-
\exp\left(-{\alpha_S\pi^2\over 
{2 \sigma(b) N_c}}x_t^2\,xG\left(x,{1\over x_t^2}\right)\right)\right)\nonumber \\
&\times& \left[\Phi_L(x_t,z,Q^2)+\Phi_T(x_t,z,Q^2)\right] \, ,
\ee
where $\Phi_{L(T)}$ is the probability of a longitudinally (transversely) 
polarized virtual photon to split into 
a quark--anti-quark pair, and the rest is the probability of that pair to 
scatter off the hadron/nucleus. Similar expressions have been used by 
several authors to reproduce the HERA data in the $Q^2=1-10$ GeV$^2$ region
~\cite{BW1GLM}. These fits, however, are not conclusive evidence for 
screening corrections since a QCD fit with appropriately adjusted parton 
distributions also reproduces this data.

A very interesting approach to small $x$ physics is through the study
of ``Onium'' scattering~\cite{Mueller94}. The large mass of the
quarkonium state provides the large scale at which the coupling
constant is evaluated. At high energies, the Onium state contains a
large number of soft gluons in addition to the quark--anti-quark pair.
In the large $N_c$ limit, these gluons can be viewed as color dipoles.
The cross--section for Onium scattering is then given by the product
of the number of color dipoles in each Onium state times the
elementary dipole--dipole scattering cross--section. The dipole
density in the Onium state obeys an integral equation whose kernel is
none other than the BFKL kernel. In this Onium picture, the scattering
cross--section grows rapidly because the number of dipoles in the
wavefunction multiplies rapidly at high energies.  The Onium formalism
thus gives us a way to quantify when multi--Pomeron effects, due to
overlapping dipoles, overtakes BFKL multiplication.

These multi--Pomeron effects are again easier to quantify in DIS off a
very large nucleus.  In this generalized case, the dipole density in the
$q\bar{q}$--pair wavefunction obeys a non--linear integral
equation~\cite{Kovchegov,Balitsky}, which sums up Pomeron ``fan''
diagrams. Pomeron loop contributions are suppressed if $\alpha_S^2
A^{1/3}$ is large. This non--linear equation has been solved
perturbatively outside the saturation region $k_t > Q_s$~\cite{yuri}.

The reader might wonder how the two approaches to high parton densities 
discussed in this section are related. It has been argued recently, that 
the non--linear integral equation for the dipole density in the Onium 
state can be obtained as a particular limit of the non--linear Wilson 
RG--equation~\cite{KMW}.

\section{Hard Diffraction}
\vskip 0.05in

With the discovery of hard diffraction by UA8~\cite{UA8}, and 
subsequent experiments at Fermilab and HERA, diffraction is again 
a hot topic. For nice recent reviews of accompanying theoretical developments, 
see Refs.~\cite{Ingelman,Hebecker}. In QCD, naively, the struck quark 
forms a colored string with the rest of the hadron--the probability of a 
gap decreases exponentially with the size of the gap. One can 
define hard diffraction as events with hard final states accompanied 
by large rapidity gaps that are not exponentially suppressed. Monte Carlo 
event generators such as POMPYT~\cite{Ingelman} which allow color singlet 
exchanges do a better job of describing the data than do event generators 
which contain only colored strings.

Diffraction has traditionally been interpreted in terms of Pomeron exchange. 
Hard diffraction is especially interesting because it lets us probe the parton 
content of the exchanged color singlet object. In a phenomenological 
picture~\cite{IngSch}, the cross section for single hard diffraction 
(to pick one of several topologies) is
\be
d\sigma({\bar p}+p\longrightarrow p +{\mbox 2 jets}) = f_{{\cp}\over p}(x_{\cp},t)
\,d\sigma (\cp + {\bar p}\longrightarrow {\mbox 2 jets}) \, ,
\ee
where the Pomeron flux factor $f_{{\cp}\over p}(x,t)\propto (1/x_{\cp})^{
2\alpha(t)-1}$. Here $\alpha(t)$ is the same function as that defined 
below Eq.~\ref{pomamp}. One can then write the factorized expression 
\be
d\sigma (\cp +{\bar p}\longrightarrow {\mbox 2 jets})=\int dx_1\,dx_2\,
dt \sum_{ji} f_{i\over \cp}(x_1,Q^2)\,f_{j\over {\bar p}}(x_2,Q^2)\,
{d\sigma_{ij\longrightarrow {\mbox 2 jets}}\over dt} \, ,
\ee
where $f_{i\over \cp}$ is the probability of finding a parton $i$ in the 
Pomeron. The CERN and Fermilab collider data seem to suggest that there is 
a larger $q\bar {q}$ than a glue component in the Pomeron, though both 
data don't seem to agree with model predictions. One problem lies with 
the difficulty in defining absolute normalizations for the Pomeron 
flux~\cite{Goulianos}. 
A more serious problem may however be the breakdown of the factorization 
hypothesis for hadron--hadron scattering.

Hard diffraction has also been studied extensively at HERA where it 
comprises $\sim 10$\% of the cross--section! In analogy to $F_2$, one can 
define an experimental observable--the diffractive structure function 
$F_2^{D(4)}(x,Q^2,x_{\cp},t)$, in terms of the differential cross--section 
for the process $ep\longrightarrow ep+X$. The more inclusive variable 
$F_2^{D(3)}(x,Q^2,x_{\cp})$ is easier to measure. Following Ingelman and 
Schlein, this can be factorized as $F_2^{D(3)} = f_{\cp \over p}(x_{\cp},t)
\,F_2^{\cp}(\beta,Q^2)$, where $\beta=x/x_{\cp}$ is the fraction of the 
Pomeron momentum carried by the parton. Recent HERA data~\cite{ZPC76} show 
deviations from universal factorization--i.e., the flux factor shows 
a $\beta$ dependence. Fits which include sub--leading Reggeon exchange 
show agreement with the Pomeron intercept $\alpha(0)\sim 1.08$ only at 
the $3\sigma$ level. The diffractive structure function shows a very weak 
dependence on $Q^2$--hard diffraction is a leading twist phenomenon.

There has been considerable theoretical work recently suggesting that, 
in exact analogy to the usual structure functions, one may define universal 
diffractive structure functions in diffractive DIS~\cite{fact}. They may 
be identified as the matrix elements of bi-local field operators, and 
shown to obey 
leading twist RG--equations. This factorization breaks down when there is 
more than one hadron in the final state (unlike DIS). The reason why it 
breaks down is that gluons from the color singlet exchange may coherently 
scatter off gluons in the other hadron--unlike inclusive structure functions, 
these processes do not cancel. This breakdown of factorization has been 
shown empirically--diffractive structure functions from HERA, 
used to compute diffractive cross-sections at the Tevatron, vastly 
overpredict the experimental data~\cite{Whitmore}.

The HERA data have been analysed within the framework of diffractive 
parton distributions~\cite{HKS}. 
It is found that the gluon component, predictably, dominates the quark 
component of the diffractive distribution. Also, the data are consistent 
with the presence of a semi--hard saturation scale. Is this scale the 
same as the scale $Q_s$ discussed earlier? It may be so since 
phenomenological models that explicitly include saturation are quite 
successful in fitting the data~\cite{BW2}. 
Such a result also arises in an approach where the 
scattering off the hadron is modelled by scattering off semi--classical 
color fields of the target using the eikonal approximation~\cite{BuHeMc}. 
In the classical EFT approach, the difference between inclusive and 
quasi--elastic 
diffractive cross--sections is simply the following~\cite{yurilarry}. 
In the former case, one squares the amplitude 
before averaging over the random color sources; in the latter, one averages 
over the amplitude with the color sources before squaring the result. The 
energy dependence of rapidity gaps has recently 
been studied in the multi--Pomeron fan diagram approach~\cite{yurigenya}. (For 
earlier related work, see for instance Ref.~\cite{BartIng}.) It will be 
interesting to see how it arises in the classical EFT approach. 

\section{Small $x$ physics and heavy ion collisions}
\vskip 0.05in

Much of the interest in heavy ion collisions have to do with the possibility 
of forming a quark gluon plasma at RHIC and LHC energies. At these 
energies, whether a plasma is formed, and how it formed, 
depends strongly on the initial conditions 
in the collision~\cite{Comments}. 
These in turn strongly depend on the small $x$ parton 
distributions in the nuclei. For momenta $k_t\sim Q_s$, coherence effects 
are significant, and the factorization picture of mini--jet production 
may break down. At what energies that happens is a quantitative question 
that has no clear answer thus far~\cite{Eskola}.

The nice thing about the classical fields 
approach is that it provides a consistent space--time picture of the 
collision~\cite{KLW}. The initial conditions are obtained by matching 
the Yang--Mills equations, in the forward and backward light cones, along 
the lightcone. We remind the reader that analytic solutions are known for 
the classical fields in the nuclei before the collision~\cite{JKMW,Kov}. 
In QCD, at small $x$, the classical $2\longrightarrow 1$ 
process dominates. Naively, in collinear factorization, this process would 
be suppressed in favor of the $2\longrightarrow 2$ process.

Gluon production in nuclear collisions is computed perturbatively, and 
is found to be infra-red divergent~\cite{KLW}. Recently, the Yang--Mills 
equations have been  solved non--perturbatively to all orders in the 
classical background field, and the energy 
and number distributions computed~\cite{AlexRaj}. One finds, 
self--consistently, a
``formation time'' beyond which it is meaningful to define these objects 
as partons--as 
opposed to field amplitudes and energies. The most relevant results
are the following. The energy distribution of gluons produced 
per unit area per unit rapidity is 
\be
{1\over \pi R^2}\, \left({dE\over d\eta}\right)_{\Delta \eta=1} = {(N_c^2-1)\over N_c}\,
{c(Q_s^2 R^2)\over 4\pi^2 \alpha_S}\, Q_s^3 \, ,
\ee
where $c(Q_s^2 R^2)\approx 4.5$, is approximately constant in the regime 
of interest for RHIC and LHC. It has been estimated that $Q_s\sim 1$ GeV 
for RHIC and $Q_s\sim 2$--$3$ GeV at LHC~\cite{Mueller99}. Similarly, 
one can compute the number per unit area per unit rapidity, and one 
finds $dN/(\pi R^2)/d\eta = {{\tilde c}\over 4\pi^2\alpha_S}\,
{(N_c^2-1)\over N_c}Q_s^2$, where ${\tilde c}\sim 1.08$. 
For $k_t\gg Q_s$, number distributions fall as $1/k_t^4$, but saturate at 
smaller values of $k_t$--the distributions are infrared finite.

How does a nuclear collision proceed from its very earliest moments? 
When produced, the gluons are on a transverse sheet, with typical momenta 
$k_t\sim Q_s$. They begin to scatter slowly (small angle scattering dominates) 
off the sheet, acquiring longitudinal momenta. This process is described 
by solving the Boltzmann equation for the single 
particle distributions~\cite{Mueller99}. The approach to equilibrium can 
be studied numerically, and the initial temperature and chemical potential 
of the equilibrated quark--gluon plasma can be extracted as a function of 
the only scale the problem--the saturation scale $Q_s$~\cite{Bjoraker}. 
Equivalently, in principle, the scale $Q_s$ can be extracted from 
studying final states in heavy--ion collisions. Besides energy and 
multiplicity distributions, rapidity correlations in event by event 
fluctuations, would also be sensitive to the saturation scale~\cite{KovLevMcL}.
If successful, heavy ion 
collisions will provide important information not only about a hot gluon 
plasma, but also about a cool color glass condensate.

\section{Outlook}
\vskip 0.05in

Our current understanding of small $x$ physics is that pQCD works at 
HERA and the Tevatron, but perhaps better than we expect it to. There is 
much flexibility in parton distributions to hide interesting new effects. 
Indeed, there are strong hints from HERA that we are on the threshold of 
a new regime of truly high parton densities, where one may expect qualitative 
changes in the behavior of distributions. 

Exciting times lie ahead. The Relativistic Heavy Ion Collider (RHIC)
will start collecting data soon, hopefully providing us with a window
to study the intial strong field strength regime in QCD and the
possible subsequent phase transition in hot and dense parton matter. 
Proposals are 
afoot to study electron DIS off nuclei at HERA energies both at DESY 
and at BNL. The latter project is now known by the acryonm eRHIC. 
{\it The 
nuclear advantage is that parton densities that would be probed only at 
c.m energies comparable to LHC c.m. energies with an $ep$ collider, 
are accessible at RHIC c.m 
energies with an $eA$ collider!} Multi--particle production is still one 
of the least understood aspects of QCD. Hopefully, the next generation of 
experiments will help us reveal its mysteries.

\noindent {\bf Acknowledgements.}

I would like to thank the organizers of Whepp-6 for their very kind 
hospitality. This work was supported by DOE Contract No. DE-AC02-98CH10886 
and by an LDRD grant at BNL.


\begin{thebibliography}{99}

\bibitem{DonnLndshff}A.~Donnachie and P.~V.~Landshoff, 
{\em Phys. Lett.}  {\bf B296} (1992) 227.

\bibitem{LowNussinov}F.~E.~Low, {\em Phys. Rev.}  {\bf D12} (1975) 163; 
S.~Nussinov, {\em Phys. Rev.}  {\bf D14}, 246 (1976).

\bibitem{BFKL}E. A. Kuraev, L. N. Lipatov and V. S. Fadin,
{\it Sov. Phys. JETP} {\bf 45}, 199 (1977); Ya. Ya. Balitsky and
L. N. Lipatov, {Sov. J. Nucl. Phys.} {\bf 28} (1978) 22.

\bibitem{Mueller94}A.~H.~Mueller and B.~Patel, 
{\em Nucl. Phys.}  {\bf B415}, 373 (1994); {\em Nucl. Phys.}
 {\bf B425}, 471 (1994).

\bibitem{Kovchegov}Y. V. Kovchegov, {\em Phys. Rev.} {\bf D60} 034008 (1999).

\bibitem{JKMW}J. Jalilian--Marian, A. Kovner, L. McLerran and H.
Weigert,
{\it Phys. Rev.} {\bf D55} (1997) 5414.

\bibitem{JKW}J. Jalilian--Marian, A. Kovner and H. Weigert, 
{\em Phys. Rev.} {\bf D59} 014015 (1999).

\bibitem{MV} L. McLerran and R. Venugopalan, {\it Phys. Rev.} {\bf D49}
(1994) 2233; {\bf D49} (1994) 3352; {\bf D50} (1994) 2225.

\bibitem{KMW}A. Kovner, J. Milhano, and H. Weigert, hep-ph/0004014.

\bibitem{Collins}J. C. Collins, {\em Phys. Rev.} {\bf D57} 3051 (1998).

\bibitem{CFS}J. C. Collins, L. Frankfurt, and M. Strikman, {\em Phys. Lett.} 
{\bf B307} 161 (1993); {\em Phys. Rev.} {\bf D56} 2982 (1997).

\bibitem{Gavin}G. P. Salam, {\em Acta Phys. Polon.} {\bf B30} 3679 (1999).

\bibitem{DGLAP} V. N. Gribov and L. N. Lipatov, {\it Sov. J. Nucl. Phys.}
{\bf 15} (1972) 78;
G. Altarelli and G. Parisi, {\it Nucl. Phys.} {\bf B126} 298
(1977); Yu. L. Dokshitzer, {\it Sov. Phys. JETP} {\bf 73} (1977) 1216.

\bibitem{GrossRujula}D. Gross, {\em Phys. Rev. Lett.} {\bf 32} 1071 (1974); 
A. DeRujula et al., {\em Phys. Rev.} {\bf D10} 1649 (1974).

\bibitem{BallForte}R. D. Ball and S. Forte, {\em Phys. Lett.} {\bf B358} 
365 (1995).

\bibitem{Forte}R. D. Ball and S. Forte, {\em Phys. Lett.} {\bf B465} 271 
(1999).

\bibitem{ManSaaWei}L.~Mankiewicz, A.~Saalfeld and T.~Weigl, 
{\em Phys. Lett.} {\bf B393} 175 (1997 ); hep-ph/9706330. 

\bibitem{McDermott}see for instance, M. McDermott, {\em Phys. Lett.} 
{\bf B349} 189 (1995).

\bibitem{FLCC}V. S. Fadin and L. N. Lipatov, {\em Phys. Lett.} 
{\bf B429}, 127 (1998);
G. Camici and M. Ciafaloni, {\it Phys. Lett.} {\bf B412}
(1997) 396;  Erratum-ibid., {\bf B417} (1998) 390.

\bibitem{fact}A. Berera and D. E. Soper, {\em Phys. Rev.} {\bf D53}, 6162 
(1996); L. Trentadue and G. Veneziano, {\em Phys. Lett.} {\bf B323} 201.

\bibitem{Whitmore}L.~Alvero, J.~C.~Collins, J.~Terron and J.~J.~Whitmore, 
{\em Phys. Rev.}  {\bf D59}, 074022 (1999).

\bibitem{Ross}D.~A.~Ross, {\em Phys. Lett.}  {\bf B431}, 161 (1998); 
E.~Levin, hep-ph/9806228.

\bibitem{CCS} 
M.~Ciafaloni, D.~Colferai and G.~P.~Salam, {\em Phys. Rev.} {\bf D60}, 
114036 (1999).

\bibitem{Mueller97}A.~H.~Mueller, 
{\em Phys. Lett.}  {\bf B396}, 251 (1997).

\bibitem{KovMuell}Y.~V.~Kovchegov and A.~H.~Mueller, 
{\em Phys. Lett.} {\bf B439}, 428 (1998).

\bibitem{RajGavai}R. V. Gavai and R. Venugopalan, {\em Phys. Rev.} {\bf D54} 
5795 (1996).

\bibitem{Kov}Y.~V.~Kovchegov, {\em Phys. Rev.} {\bf D54}, 5463 (1996).

\bibitem{AJMV}A. Ayala, J. Jalilian--Marian, L. McLerran and R.
Venugopalan,
{\it Phys.Rev.} {\bf D52} (1995) 2935; {\it ibid.} {\bf  D53} (1996) 458.

\bibitem{JKLW1}J. Jalilian--Marian, A. Kovner, A. Leonidov and H.
Weigert,
{\it Nucl. Phys.} {\bf B504} (1997) 415.

\bibitem{JKLW2}J. Jalilian-Marian, A. Kovner, A. Leonidov and H. Weigert, 
{\em Phys. Rev.} {\bf D59}, 034007 (1999).

\bibitem{GLR}
L.~V.~Gribov, E.~M.~Levin and M.~G.~Ryskin, {\em Phys. Rept.} 
{\bf 100}, 1 (1983).

\bibitem{MuellQiu}A.~H.~Mueller and J.~Qiu, 
{\em Nucl. Phys.} {\bf B268}, 427 (1986).

\bibitem{MV99}L. McLerran and R. Venugopalan, {\em Phys. Rev.} {\bf D59} 
(1999) 094002; R.~Venugopalan, {Acta Phys. Polon.}  {\bf B30}, 3731 (1999).

\bibitem{M90NZ}
A. H. Mueller, {\em Nucl. Phys.} {\bf B335} (1990) 115; 
N. N. Nikolaev and B. G. Zakharov, {\em Z. Phys.} {\bf C49} 
(1991) 607.

\bibitem{BW1GLM}E. Gotsman, E. Levin, U. Maor, and E. Naftali, {\em Nucl.
Phys.} {\bf B539}, 535 (1999); {\em Phys. Lett.} {\bf B425} (1998) 36; 
K. Golec-Biernat, M. W\"{u}sthoff,   
{\em Phys. Rev.} {\bf D59},  014017 (1999); A.L. Ayala Filho, 
M.B. Gay Ducati, and E.M. Levin {\em Eur.Phys.J.} {\bf C8}115 (1999). 
A. L. Ayala Filho, M. B. Gay Ducati, and V. P. Gonsalves, {\em Phys. Rev.} 
{\bf D59} 054010 (1999). 

\bibitem{Balitsky}I. Balitsky, {\em Phys. Rev.} {\bf D60} 014020 (1999); 
{\em Phys. Rev. Lett.} {\bf 81} 2024 (1998).

\bibitem{yuri}Y.~V.~Kovchegov, 
{\em Phys. Rev.}  {\bf D61}, 074018 (2000); E.~Levin and K.~Tuchin, 
hep-ph/9908317.

\bibitem{UA8}R. Bonino et al., {\em Phys. Lett.} {\bf B211} 239 (1988).

\bibitem{Ingelman}G.~Ingelman, hep-ph/9912534.

\bibitem{Hebecker}A.~Hebecker, {\em Acta Phys. Polon.}  {\bf B30}, 3777 (1999).

\bibitem{IngSch}G. Ingelman and P. Schlein, {\em Phys. Lett.} {\bf B152} 
256 (1985).

\bibitem{Goulianos}
K.~Goulianos and J.~Montanha, {\em Phys. Rev.}  {\bf D59}, 114017 (1999).

\bibitem{ZPC76}C. Adloff et al., {\em Z. Phys.} {C76} 613 (1997); 
J. Breitwig et al., {\em Eur. Phys. J.} {\bf C6} 43 (1999).

\bibitem{HKS}F.~Hautmann, Z.~Kunszt and D.~E.~Soper, 
{\em Nucl. Phys.}  {\bf B563}, 153 (1999).

\bibitem{BW2}K. Golec-Biernat, M. W\"{u}sthoff, {\em Phys. Rev.}  
{\bf D60}, 114023 (1999).

\bibitem{BuHeMc}W. Buchm\"uller and A. Hebecker, {\em Nucl. Phys.} {\bf B476} 
203 (1996); W. Buchm\"uller, M. McDermott, and A. Hebecker, {\em Nucl. Phys.} 
{\bf B500} 621 (1997).

\bibitem{yurilarry}Y.~V.~Kovchegov and L.~McLerran, 
{\em Phys. Rev.}  {\bf D60}, 054025 (1999).

\bibitem{yurigenya}Y.~V.~Kovchegov and E.~Levin, hep-ph/9911523.

\bibitem{BartIng}J. Bartels and G. Ingelman, {\em Phys. Lett.} {\bf B235} 
(175) (1990).

\bibitem{Comments}
R.~Venugopalan, {\em Comments Nucl. Part. Phys.}  {\bf 22}, 113 (1997).

\bibitem{Eskola}K.~J.~Eskola, {\em Comments Nucl. Part. Phys.} 
{\bf 22}, 185 (1998).

\bibitem{KLW}
A. Kovner, L. McLerran and H. Weigert, 
{\em Phys. Rev} {\bf D52} 3809 (1995); {\bf D52} 6231 (1995); 
Yuri Kovchegov and Dirk Rischke, {\em Phys. Rev.} {\bf C56}
1084 (1997); S. G. Matinyan, B. M\"uller and D. H. Rischke,
{\em Phys. Rev.} {\bf C56} (1997) 2191;M. Gyulassy and L. McLerran, 
{\em Phys. Rev.} {\bf C56} (1997) 2219; Xiao-feng Guo, {\em Phys. Rev.} {\bf D59} 094017 (1999).

\bibitem{AlexRaj}
A.~Krasnitz and R.~Venugopalan, hep-ph/0004116; hep-ph/9909203, to appear in 
{\em Phys. Rev. Lett.}; {\em Nucl. Phys.}  {\bf B557}, 237 (1999); 
hep-ph/9808332, hep-ph/9706329. 

\bibitem{Mueller99}A.~H.~Mueller, {\em Phys. Lett.}  {\bf B475}, 220 (2000); 
A.~H.~Mueller, hep-ph/9906322.

\bibitem{Bjoraker}J. Bjoraker and R. Venugopalan, in preparation.

\bibitem{KovLevMcL}Y.~V.~Kovchegov, E.~Levin and L.~McLerran, hep-ph/9912367.

\end{thebibliography}

\end{document}